\documentclass[reprint,amsmath,amssymb,
prapplied,longbibliography,
]{revtex4-2}
\usepackage{comment}
\usepackage{xcolor, soul}
\usepackage{graphicx}
\usepackage{subfigure}
\usepackage{dcolumn}
\usepackage{bm}
\usepackage{multirow}
\usepackage{amsmath}
 
\usepackage{braket}
\usepackage{diagbox, eqparbox, hhline}
\usepackage{hyperref}
\usepackage{tikz-cd}
\usepackage{soul}
\tikzcdset{every label/.append style = {font = \normalsize}}

\DeclareMathAlphabet{\mathpzc}{OT1}{pzc}{m}{it}

\usepackage{graphicx}
\usepackage{dcolumn}
\usepackage{bm}

\usepackage[utf8]{inputenc}
\usepackage[T1]{fontenc}

\begin{document}

\title{Harnessing the 
superconducting diode effect through inhomogeneous magnetic fields}
\author{Leonardo Rodrigues Cadorim}
\author{Edson Sardella}%
\affiliation{Departamento de F\'isica, Faculdade de Ci\^encias, 
Universidade Estadual Paulista (UNESP), Caixa Postal 473, 
17033-360, Bauru-SP, Brazil}%
\author{Cl\'ecio C.\ de Souza Silva}
\affiliation{
Departamento de Física, Centro de Ciências Exatas e da Natureza, 
Universidade Federal de Pernambuco, Recife–PE, 50670-901, Brazil
}
\email{Corresponding author: clecio.cssilva@ufpe.br}

\date{\today}

\begin{abstract}
We propose a superconducting diode device comprising a central superconducting film flanked by two wires 
carrying an applied DC bias, suitably chosen so as to generate different asymmetric field profiles. Through numerical simulations of the coupled Ginzburg-Landau and heat-diffusion equations, we show that this design 
is capable of efficiently breaking the reciprocity of the critical current in the central superconductor, thus promoting the diode effect in response to an applied AC current. By adjusting the DC bias in the wires, we find the optimum inhomogeneous field profile 
that facilitates the entrance of vortices and antivortices in a given polarity of the applied AC current and impede their entrance in the other polarity. This way, the system behaves as a superconducting half-wave rectifier, with diode efficiencies surpassing 70\%. Furthermore, we detail the behavior and diode efficiency of the system under different experimental conditions, such as the substrate heat transfer coefficient and the sweep rate of the external current.
\end{abstract}

\maketitle

%

\section{\label{sec:level1}Introduction}

In modern electronics, the search for faster and efficient devices is a permanent practice aiming for enhanced performance. 
In the midst of this search, the superconducting diode 
has played a prominent role \cite{moll2023,nadeem2023}. Such devices combine all the functionality of semiconductor diodes, \textit{e.g.}, the conversion of alternate (AC) currents into direct currents (DC), with the vantage of carrying electricity without resistance.

At the core of the superconducting diode lies a fundamental characteristic, the absence of symmetry in current flow, expressed as $I_{c}^+ \neq I_{c}^-$, where $I_c^{\pm}$ is the critical current leading to the resistive state at a given current direction. 
However, the superconducting diode concept originated decades ago as a mean of manipulating the density and motion of magnetic flux quanta, known as vortices, within superconductors~\cite{lee1999reducing,olson2001}. The primary goal was to disrupt the inversion symmetry of the pinning potential acting on vortices. This would enable an alternating current to induce vortex motion solely along the potential's easy direction, resulting in a resistive flux-flow state during one half cycle and purely superconducting current flow during the other one. Subsequently, experimental realization of these flux quanta diodes swiftly followed, with diverse research groups employing various nanolithographic techniques to selectively break spatial inversion symmetry for vortex motion, such as regular arrays of asymmetric dots and antidots~\cite{villegas2003superconducting,vandeVondel2005vortex,deSouzaSilva2006Nature}, asymmetric distributions of symmetric dots and antidots~\cite{wordenweber2004guidance,Gillijns2007,deLara2010vortex}, regular arrays of nanosized magnetic dipoles~\cite{deSouzaSilva2007dipole,silhanek2007manipulation}, and asymmetric weak-pinning channels~\cite{yu2007}.

Recently, superconducting diodes have gained renewed interest, fueled in part by the discovery of novel intrinsic mechanisms governing nonreciprocal current flow. This resurgence encompasses diverse systems, ranging from heterostructures composed of superconducting materials and topological insulators or ferromagnetic materials \cite{yasuda2019,narita2022,hou2023,legg2022,mehrnejat2023,karabassov2023,gutfreund2023}, different superconducting systems with an applied magnetic field \cite{wakatsuki2017,hoshino2018,ando2020,zhang2020,bauriedl2022,margineda2023,chahid2023,yuan2022}, and even superconducting systems with no external field~\cite{golod2022,wu2022,lin2022,jiang2022}. Yet another possibility comes from Josephson junction based systems, where the diode effect is achieved due to the asymmetric current-phase relation of the junction \cite{ghosh2024,steiner2023,zazunov2024,cheng2023,zinkl2022,gupta2023}.
This diverse exploration marks an intriguing phase in the study of superconducting diodes and their various applications.

Intrinsic superconducting diodes  
rely on non-reciprocal pair-breaking mechanisms inherent to the material~\cite{Daido2022,Ilic2022,Banerjee2024}. Although these mechanisms are of fundamental importance, they lack the high degree of controllability of vortex-based red superconducting diodes, 
where the non-reciprocal critical currents can be carefully designed by e.g. nanostructuring. Furthermore, in typical experimental situations, the superconducting sample enters the resistive state at currents considerably lower than the theoretical depairing current, either as a result of vortex motion or the nucleation of supercritical hot spots. This makes it challenging to discern experimentally intrinsic and non-intrinsic effects~\cite{qiao2023}. 


During the resistive state of a vortex-based diode, the material is still superconducting and dissipation is induced by vortex motion (flux-flow). Since the flux-flow resistance is necessarily lower than the normal state resistance, the measured DC signals in this case are relatively small. This is a clear disadvantage when compared to diodes where the dissipationless superconducting state for current applied in one direction transitions directly to the fully normal state when the current is reversed, leading to a considerably larger DC signal. We shall refer to these two scenarios as flux-flow dominated (FFD) and normal-state dominated (NSD) diode effects, respectively.
Remarkably, Hou {\it et al.}~\cite{hou2023} recently reported on the observation of highly efficient diode effect on conventional superconducting films where the AC switching between superconducting and normal states is induced by non-reciprocal penetration of vortices due to asymmetrical conditions at the sample edges, that is, by an extrinsic, vortex-based mechanism. However, the reason why the flux-flow state was suppressed in favor of a sudden jump to the full normal state, thus making this FFD diode into a NSD diode, is unclear. A possible explanation can be inferred from Ref.~\cite{lyu2021superconducting}, where the authors investigated the diode effect in a superconducting stripe with a conformal array of nanoholes of broken inversion symmetry. This work suggests that hot posts spots induced by fast vortex motion can drive the superconductor directly into the normal state thus bridging the gap between FFD and NSD scenarios.

These recent advances emphasize the necessity of better understanding the role of heating effects in superconducting diodes. Additionally, the suggestion that a FFD diode can be made into a NSD diode redirects the research on vortex diode systems towards the optimization of the diode efficiency, defined as $\epsilon=|I_c^--I_c^+|/(I_c^-+I_c^+)$. Tackling these challenges is a key point for obtaining more efficient and reliable superconducting diodes.

\begin{figure}[!t]
    \centering
    \includegraphics[width=\columnwidth]{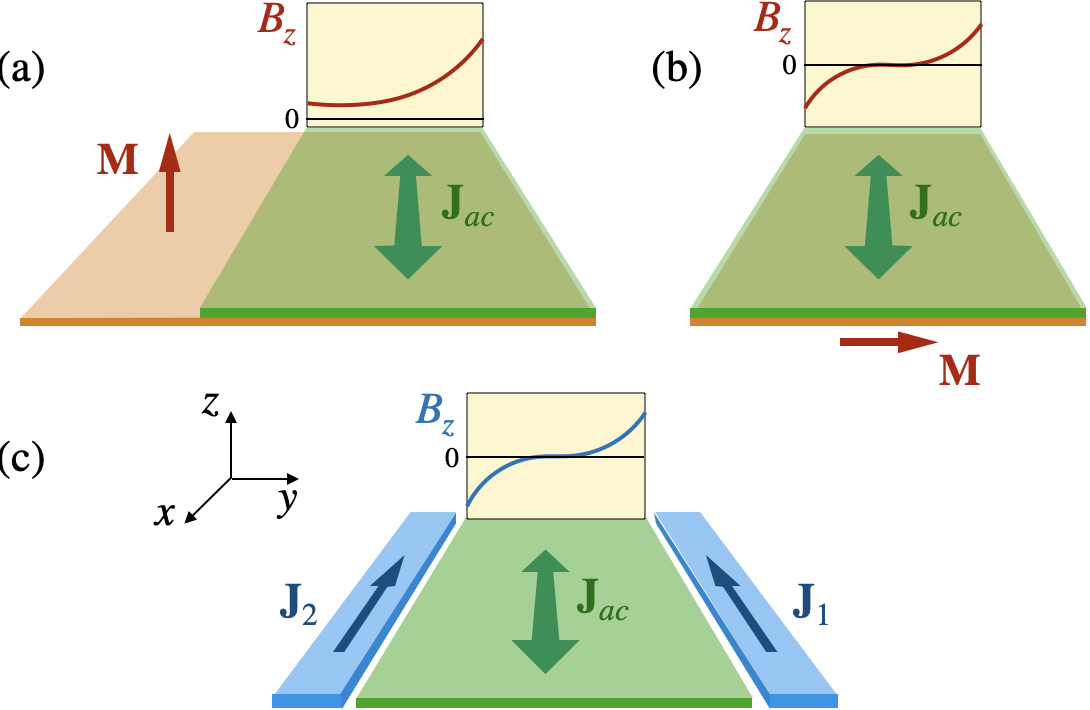}
    \caption{Schematic view of a superconducting film (green) subjected to asymmetric field profiles $B_z(y)$ induced by: (a) an asymmetrically lying ferromagnetic film (orange) with off-plane magnetization ${\bf M}$; (b) the same as (a) but with in-plane magnetization and symmetric arrangement of the bilayer; and (c) currents applied onto lateral superconducting stripes (blue). ${\bf J}_1$ and ${\bf J}_2$ can be adjusted to generate different field profiles. Here, setting ${\bf J}_2={\bf J}_1=-J\hat{\bf x}$ emulates $B_z(y)$ similar to that induced by the magnet in (b). In all cases, an alternating current ${\bf J}_{ac}$ applied parallel to $x$ induces nonreciprocal vortex penetration and motion.
    }
    \label{fig:fig1}
\end{figure}

In the present work, we investigate the interplay between the flux-flow dominated (FFD) and normal-state dominated (NSD) operation modes of the superconducting diode effect in thin superconducting films exposed to external asymmetric fields. By integrating the coupled time-dependent Ginzburg-Landau and heat diffusion equations, we explore simple inhomogeneous field landscapes, from asymmetric fully positive to antisymmetric (half-positive, half-negative) flux profiles, induced either by ferromagnetic films or DC currents in nearby wires as illustrated in Fig.~\ref{fig:fig1}. Our findings reveal that both profile types promote the FFD mode, with the antisymmetric one exhibiting the highest diode efficiencies, up to 70\%. Furthermore, at large enough AC amplitudes, FFD is replaced by NSD as a result of vortex-antivortex collisions and subsequent hotspot proliferation. We further demonstrate that reducing the substrate's heat transfer coefficient accelerates the FFD-to-NSD transition, thereby enhancing the power output of the diode. These insights illuminate recent discoveries of highly efficient superconducting diodes and emphasize the role of vortex-antivortex collisions and the importance of optimizing substrate designs and cooling environments for future superconducting diode platforms.
%
%

The outline of this work is as follows. In Sec.~\ref{sec:level2} we present our system and the theoretical formalism used to investigate it. Our results and their discussion are presented in Sec.~\ref{sec:level3}. Finally, we present our concluding remarks in Sec.~\ref{sec:level4}.

\section{\label{sec:level2}Theoretical Formalism}

The resistive state of superconductors in the presence of 
a driving current can be described by the time-dependent 
Ginzburg-Landau equations. In reduced units, these equations 
can be written as: \cite{kramer1978theory,watts1981nonequilibrium}
\begin{eqnarray}
    \frac{u}{\sqrt{1+\gamma^2|\psi|^2}}\left [ \frac{\partial }{\partial t}
    +\frac{1}{2}\gamma^2\frac{\partial |\psi|^2}{\partial t} \right ] \psi =  
    & & \nonumber \\
    \left(\mbox{\boldmath $\nabla$}-i\textbf{A}\right)^2\psi 
    +\psi(1-T-|\psi|^2), & &
    \label{eq:eq1}
\end{eqnarray}
\begin{equation}
    \sigma\frac{\partial \textbf{A}}{\partial t}
    = \textrm{Im}\left [\bar{\psi}{(\mbox{\boldmath $\nabla$}-i\textbf{A})}\psi\right]
    -\kappa^2\mbox{\boldmath $\nabla$}\times\mbox{\boldmath $\nabla$}\times\textbf{A},      
    \label{eq:eq2}
\end{equation}
where $\psi$ is the superconducting order parameter, $\textbf{A}$ is the vector potential, $\sigma$ is the normal conductivity, and $\kappa=\lambda(0)/\xi(0)$, with $\lambda(0)$ the London penetration depth and $\xi(0)$ the coherence length at zero temperature. Here we fix $u=5.79$, and $\gamma = 20$. In Eqs.~(\ref{eq:eq1}) and (\ref{eq:eq2}) we express lengths in units of $\xi(0)$, $T$ in units of the critical temperature, $T_c$, time in units of  $t_{GL}=\pi\hbar/8uk_BT_c$, $\psi$ in units of the field-free order parameter at $T=0$, $\psi_\infty(0)$, and $\textbf{A}$ in units of $\xi(0)H_{c2}(0)$, with $H_{c2}(0)$ the upper critical field at $T=0$. 


To account for the dissipation effects caused e.g. by v-av collisions, Eqs.~(\ref{eq:eq1}) and (\ref{eq:eq2}) must 
be coupled with the heat diffusion equation: 
\cite{vodolazov2005,duarte2017dynamics}
\begin{equation}
    \nu\frac{\partial T}{\partial t} = 
    \zeta\nabla^2T+\sigma\left (\frac{\partial \textbf{A}}{\partial t} \right )^2
    -\eta(T-T_0),\label{eq:eq3}
\end{equation}
where $\nu$, $\zeta$, and $\eta$ are the thermal capacity, thermal conductivity of the material, and the heat transfer coefficient of the substrate, respectively, and $T_0$ is the temperature of the thermal bath. In this work, we have used $\nu = 0.03$, $\zeta = 0.06$, and varied $\eta$ from $2.0\times 10^{-5}$ to $2.0\times 10^{-3}$.

The inhomogeneous field that allows the emergence of the diode effect can be introduced through different routes. Panels $(a)$ and $(b)$ of Fig.~\ref{fig:fig1} show two possible scenarios using a superconductor/ferromagnet heterostructure, where the inhomogeneous field comes from the ferromagnet. The system depicted in panel $(b)$ was investigated in Ref.~\onlinecite{hou2023}. In the present work, 
we introduce a new possible configuration $(c)$ comprising a central superconducting film carrying the AC current and flanked by two superconducting wires carrying DC currents suitably chosen to produce the desired inhomogeneous field profile. 
Configurations depicted in panels $(b)$ and$(c)$ have the advantage that the AC current self-field matches the profile of the inhomogeneous field for one AC current polarity and opposes it for the other one. As we shall see later on, this scenario increases the diode efficiency. We focus our detailed investigation on configuration (c), due to its versatility in producing different profiles, from fully positive to antisymmetric, by simply adjusting the lateral currents $J_1$ and $J_2$. In a practical perspective, this configuration offers the possibility of easily reversing the diode effect by changing the polarity of the side currents.

In what follows, the side stripes have width $w = 50\xi(0)$, thickness $d_s = 5\xi(0)$ and their separation from the edges of the central superconductor is $s = 5\xi(0)$. As it follows from elementary magnetostatics, the $z$ component of the total field produced by the stripes is:
\begin{eqnarray}
    H_{nh}(y) &=& \frac{J_1 d_s}{2\pi\kappa^2}\ln\left(\frac{-y+w+s+L_y/2}{-y+s+L_y/2}\right) \nonumber \\
    &-&\frac{J_2 d_s}{2\pi\kappa^2}\ln\left(\frac{y+w+s+L_y/2}{y+s+L_y/2}\right) \label{eq:eq4}
\end{eqnarray}
The central superconductor has width $L_y = 200\xi(0)$, thickness $d=\xi(0)$, and is driven by an applied AC current given by $J_{ac}(t) = J_a \sin(2\pi t/\tau)$, where $\tau$ is the period of each oscillation. 

Eqs.~(\ref{eq:eq1})-(\ref{eq:eq3}) are solved numerically only for the central superconductor, supplemented by periodic boundary conditions along $x$, with period $L_x = 400\xi(0)$.
In the Supplementary Material, we show that our choice of $L_x$ does not affect the physical behavior of the system.
At the edges $y = \pm L_y/2$, $\psi$ and $T$ satisfy the boundary conditions $ \hat{\textbf{y}}\cdot(\mbox{\boldmath $\nabla$}-i\textbf{A})\psi = 0$ and $\hat{\textbf{y}}\cdot\mbox{\boldmath $\nabla$}T=0$ \cite{vodolazov2005,duarte2017dynamics}. Additionally, the local field $\textbf{h} = \bm{\nabla} \times \textbf{A}$ at the sample edges satisfies $h_z(x,\pm L_y/2) = H_{nh}(\pm L_y/2)\pm L_yJ(t)/2\kappa^2$, which formally accounts for the applied current $J(t)$ in the sample~\cite{cadorim2020ultra}.





\section{\label{sec:level3}Results and Discussion}

To analyze the conditions for the occurrence of the superconducting diode effect or the vortex diode effect in our system, we have varied both the amplitude of the applied AC current and the values of $J_1$ and $J_2$ carried by each side stripe. We limit $J_1$ and $J_2$ to be lower than $0.52~\sigma\hbar/2e\xi(0)t_{GL}$. With the thermal bath temperature fixed at $T_0=0.96T_c$ \cite{Note}, such currents would destroy superconductivity in the central stripe. Therefore, the side stripes must be composed of a superconducting material with larger critical temperature and depairing current density. In what follows, voltages are expressed in units of $V_0=\hbar/2et_{GL}$ and current densities in units of $j_0=\sigma V_0/\xi(0)$. Further, we fix $\kappa=5$ and, unless stated otherwise, the period of the AC current is $\tau = \tau_0 = 10^5 t_{GL}$.

\begin{figure}[!b]
    \centering
    \includegraphics[width=0.99\columnwidth]{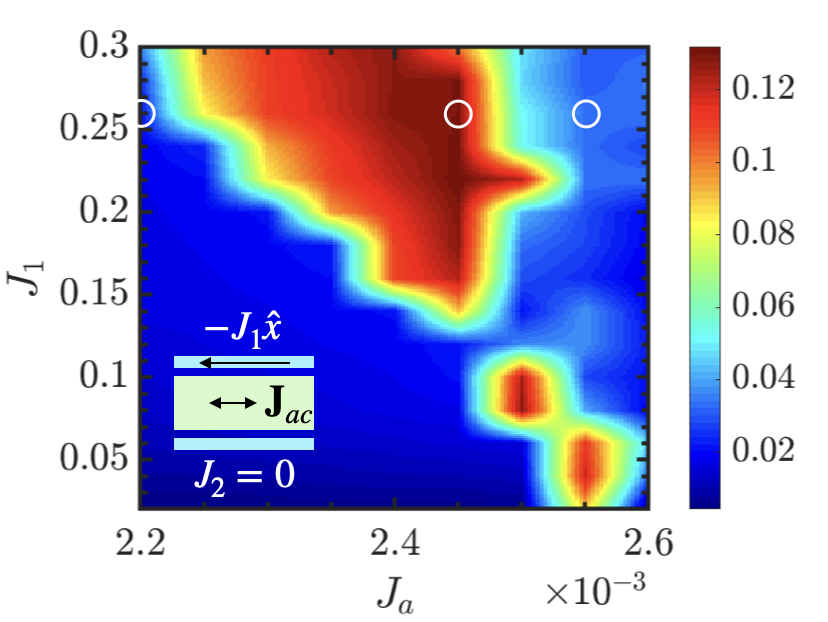}
    \caption{Phase-diagram displaying the color plot of the output voltage $V_{DC}$ (in units of $V_0=\hbar/2et_{GL}$) as a function of $J_1$ and the amplitude of the AC current (both in units of $j_0 = \sigma V_0/\xi(0)$). We fix $J_2 = 0$. }
    \label{fig:fig2}
\end{figure}

    \subsection{Fully positive asymmetric flux profiles \\($J_1 \neq 0$, $J_2 = 0$)}
    
    We start our analysis by considering the case where the asymmetric flux profile is positive everywhere in the superconducting film. This situation is accomplished by choosing $J_1>0$ and $J_2=0$ in Eq.~\eqref{eq:eq4}.
    Fig.~\ref{fig:fig2} shows the color plot of the time averaged voltage $V_{DC}$ as a function of the amplitude of the AC current and the value of $J_1$. As can be seen, if we increase the amplitude of the applied current with a fixed value of $J_1$, $V_{DC}$ starts from very small values and then gradually increases until it jumps to a region with a larger value (red region in the figure). Inside this region, the voltage continues to increase with the current amplitude. When $V_{DC}$ reaches its maximum value, the voltage output suddenly drops to another low value region and then gradually decreases as the current is further increased. In the large voltage region, $V_{DC}$ is almost ten times larger than the values obtained outside this domain. The appearance of two separate islands of $V_{DC}$ at small $J_1$ is probably an artifact of the finite step used to vary $J_a$, which might hinder finer details.
    
    \begin{figure}[!t]
        \centering
        \includegraphics[width=0.9\columnwidth]{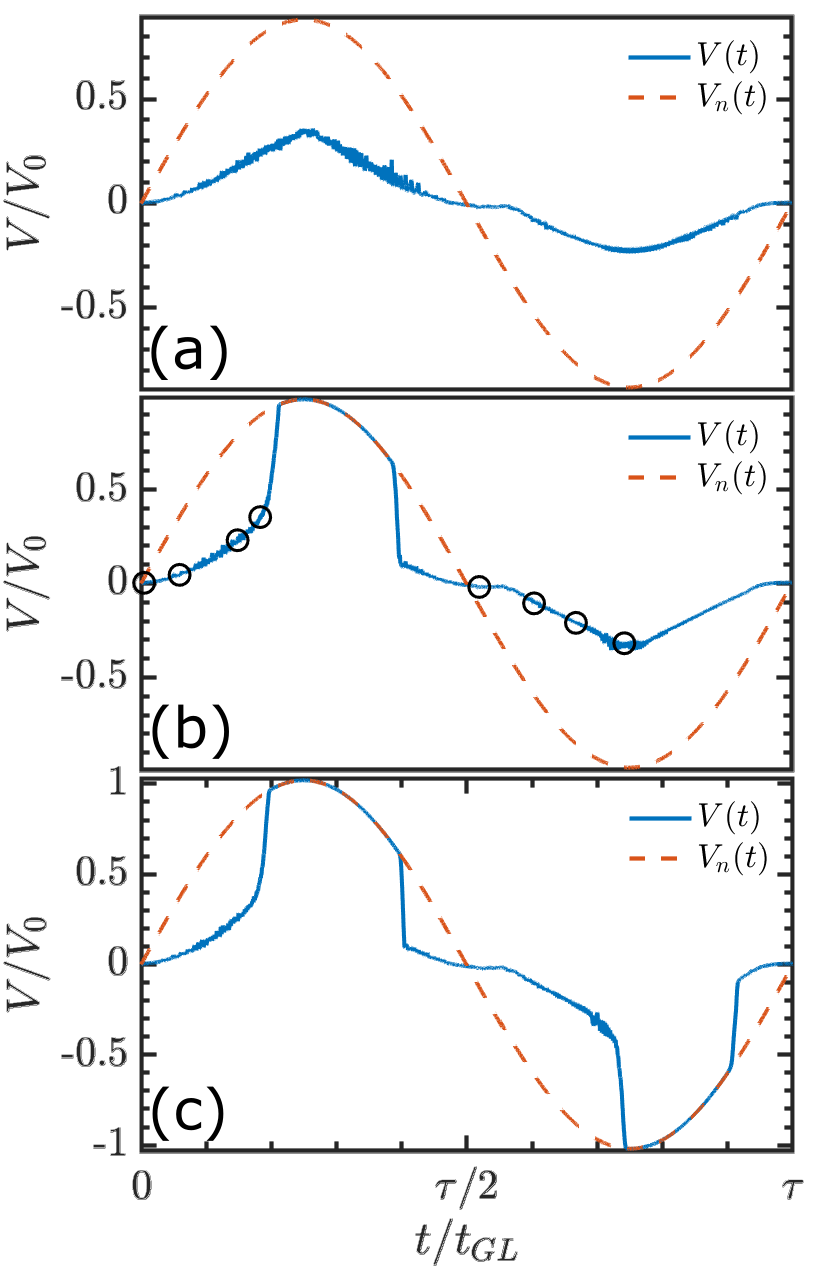}
        \caption{(Color online) The voltage signal as a function of time (solid blue line) for $J_1 = 0.26$ and three current values, $J_a = 2.20\times 10^{-3}$ (panel $(a)$), $J_a = 2.45\times 10^{-3}$ (panel $(b)$) and $J_a = 2.55\times 10^{-3}$ (panel $(c)$). In each panel, the red dashed lines represent the voltage if the system were in the normal state.}
        \label{fig:fig3}
    \end{figure}

    Fig.~\ref{fig:fig3} shows the voltage output (solid blue line) as a function of time for three different AC current amplitudes and $J_1 = 0.26$. In each panel, the dashed red line shows the voltage that would be obtained if the system were completely normal, \textit{i.e.}, $V_n(t) = (J_aL_x/\sigma)\sin\left(2\pi t/\tau\right)$.
    Panel $(a)$ corresponds to a scenario in the low $V_{DC}$ region which occurs before the region with maximum voltage. In this case, the vortex dynamics caused by the current is never sufficient to send the system to the normal state and so it stays superconducting throughout the whole period, although a finite resistance appears due to the vortex motion. Therefore, the system displays a small but finite $V_{DC}$. 
    This occurs due to the presence of the inhomogeneous field, which is positive throughout the whole width of the superconductor, but is stronger at the top edge and weaker at the bottom one. In the absence of the AC current, the inhomogeneous field nucleates vortices near the top edge of the film. For positive AC polarity, the self-field profile, that is, the field produced by the current itself, is positive at the top edge and negative at the bottom one, thus matching the external field profile where it is stronger and opposing it at its weaker side. In contrast, for the negative AC current polarity, the AC current self-field opposes the inhomogeneous field at the top edge and matches it at the bottom one. This discrepancy leads to an asymmetric vortex dynamics for each current polarity, thus leading to the finite $V_{DC}$.
    
    In panel $(b)$, we show the voltage as a function of time for a case inside the large $V_{DC}$ region. The reason for the jump in the output voltage now becomes clear. For the positive polarity of the AC current, the vortex dynamics produces enough heat to completely destroy the superconducting state for a certain interval of time. A similar behavior was found in Ref.~\onlinecite{lyu2021superconducting}.
    Superconductivity is recovered as the magnitude of the current decreases and, for the negative polarity, the normal state is never achieved. The complete destruction of superconductivity only for one polarity causes a large discrepancy between the calculated voltage in the two halves of the period, resulting then in a large $V_{DC}$. The reason for the asymmetry between positive and negative polarities will be discussed below.
    
    Finally, panel $(c)$ shows the voltage for a case of low $V_{DC}$ values that occurs after the large $V_{DC}$ region. Here, as we have also encountered for panel $(b)$, the system goes to the normal state for the positive polarity. The difference now is that the current is sufficiently strong to completely destroy the superconducting state in both polarities. As a result, the asymmetry between the two halves of the period decreases, so decreasing $V_{DC}$. Nevertheless, the average voltage continues to be finite, once the time lapse that the system remains at the normal state is larger for the positive polarity. Animations of the evolution of the order parameter for the three cases presented in Fig.~\ref{fig:fig3} can be found as Supplementary Material.

    As can be seen, the region of the phase-diagram depicted by panel $(a)$ of Fig.~\ref{fig:fig3} presents the characteristic behavior of the vortex diode effect, with the DC voltage arising from the asymmetric vortex motion for each current polarity. 
    In contrast, in panel (c), $V_{DC}$ comes from the 
    superconducting diode effect, with the major difference that the superconducting state is completely destroyed for both current polarities inside the region with enhanced output voltage.
    
    \begin{figure}[!t]
       \centering
       \includegraphics[width=0.99\columnwidth]{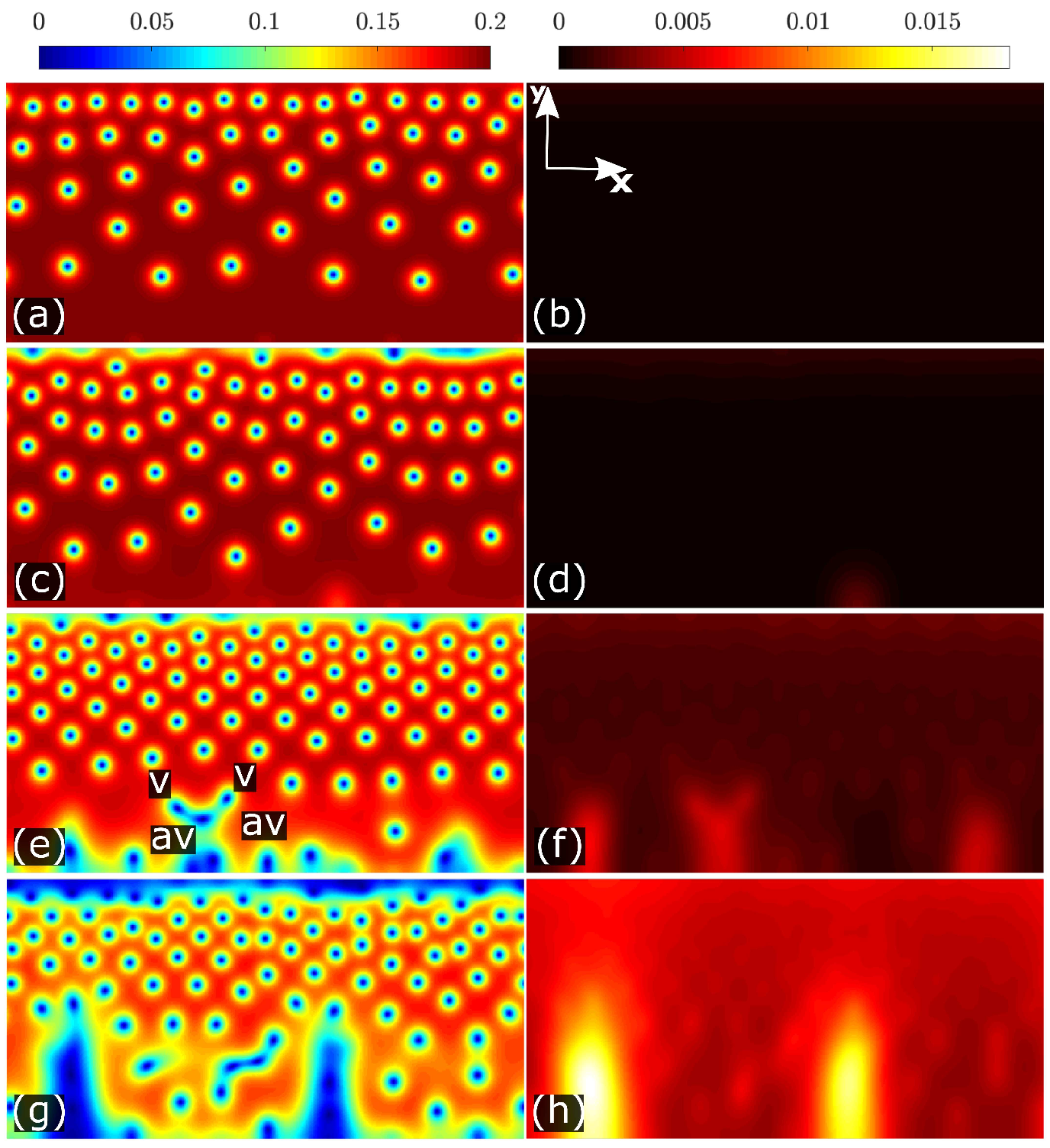}
       \caption{(Color online) Color plot of the order parameter and local temperature difference $T-T_0$ at four different times for parameters corresponding to panel $(b)$ in Fig.~\ref{fig:fig3}. Each line of the figure corresponds to a black circle in the positive current branch.}
       \label{fig:fig4}
    \end{figure}

\begin{figure}[!t]
        \centering
        \includegraphics[width=0.99\columnwidth]{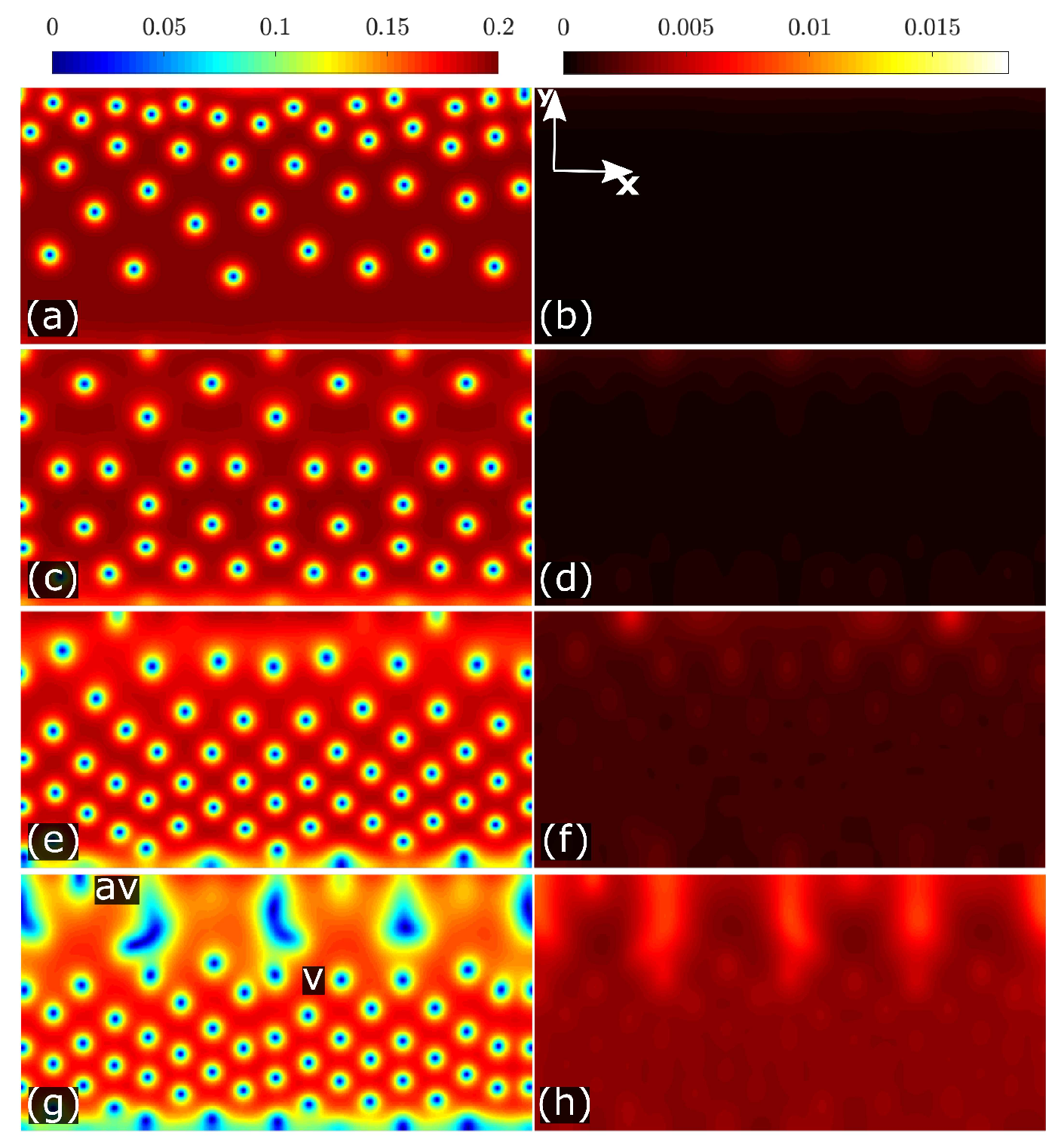}
        \caption{(Color online) Color plot of the order parameter and local temperature difference $T-T_0$ at four different times for parameters corresponding to panel $(b)$ in Fig.~\ref{fig:fig3}. Each line of the figure corresponds to a black circle in the negative current branch.}
        \label{fig:fig5}
    \end{figure}

    Once panel $(b)$ represents the more interesting case of the superconducting diode effect with enhanced $V_{DC}$, let us take a closer look at the vortex dynamics that leads to this physics. Figs. \ref{fig:fig4} and \ref{fig:fig5} depict the time evolution of the order parameter and local temperature for positive and negative AC current polarity, respectively. From top to bottom, the panels follow a sequence of time instants identified by black circles in Fig.~\ref{fig:fig3}$(b)$. 
    Panels $(a), (b)$ in Fig. \ref{fig:fig4} shows the vortex configuration at $t = 0$, 
    where the applied current vanishes. As we can see, vortices penetrate the superconductor at the top edge, 
    near the lateral stripe with nonzero current. The configuration we encountered resembles the characteristic pattern of a conformal vortex crystal~\cite{menezes2017c,menezes2019}. As the time evolves and the applied AC current becomes 
    positive, the vortices begin to move towards the opposite side of the sample $(c), (d)$, 
    at the same time new vortices nucleate at the top edge. 
    The nucleation and annihilation of vortices dissipate energy and thus locally increase the system temperature (see the animations for the evolution of the local temperature in the Supplementary Material). This local heating leads to the partial suppression of the superconducting state, but, for this small vortex velocity, the heat is quickly removed from the film and the system remains at the superconducting state.

    As the current is further increased, the vortex velocity
    increases and, simultaneously, the magnitude of the AC current becomes sufficiently large to allow the penetration of antivortices in the opposite side of the superconductor, as depicted in panels $(e)$ and $(f)$. 
    The vortices and antivortices annihilate in pairs near the bottom edge of the superconductor, in a process that dissipates enough heat to induce hot spots. 
    In this panel, to help the dynamics visualization, a few vortices and antivortices are indicated by the labels $v$ and $av$, respectively.
    Finally, for a sufficiently large magnitude of the AC current (see panels $(g)$ and $(h)$), 
    the nucleation and annihilation of vortex pairs become more frequent and the hot spots are able to spread across the sample. This ultimately leads to the complete suppression of the superconducting state, which is only reestablished at a later time after the applied current is decreased. 

    The vortex dynamics for the negative current branch is rather different. 
    %
    %
    %
    At $t = \tau/2$, the vortices again arrange themselves in the conformal crystal like pattern, as shown in  Fig.\ref{fig:fig5}, panels  $(a)$ and $(b)$. At later times, the magnitude of the applied current increases and the vortices at the top exit the sample (see panels $(c)$ and $(d)$), but at a lower rate than in the case depicted in Fig.~\ref{fig:fig4}, once the top edge displays a larger surface barrier. Simultaneously, a few vortices enter the superconductor through the bottom edge. Though both of these processes are enhanced in panels $(e)$ and $(f)$, the lower field at the bottom edge sustains a lower vortex density than in the positive branch. Since the dissipated power is proportional to the number of moving vortices and antivortices, hot spots are considerably weaker in this case.  
    This is further confirmed in panels $(g)$ and $(h)$, where antivortices nucleate 
    at the top of the superconductor and annihilation of vortex pairs occurs just as in the positive current branch. However, the dissipation in this case is never sufficient to suppress the superconducting state and the system never transitions to the normal state.
    
    \subsection{Antisymmetric flux profiles ($J_1=J_2$)} 
    
    Although the scenario depicted in Figs.~\ref{fig:fig4} and \ref{fig:fig5} produces a large output voltage $V_{DC}$, it suffers from the problem of a finite voltage in the negative polarity branch of the AC current induced by flux-flow. This nonzero resistance is a shortcoming to the very purpose of a superconducting diode, \textit{i.e.}, to have all the functionalities of a regular diode with the benefit of dissipationless transport of current in half-wave of the AC signal. A possible way to overcome this issue is to introduce a finite current $J_2$ in the second stripe, such that the surface barrier to vortex entry through the bottom edge increases up to a point where it can not be surpassed by the self-field of the AC current during the negative half cycle. 
    
    \begin{figure}[!t]
        \centering
        \includegraphics[width=0.99\columnwidth]{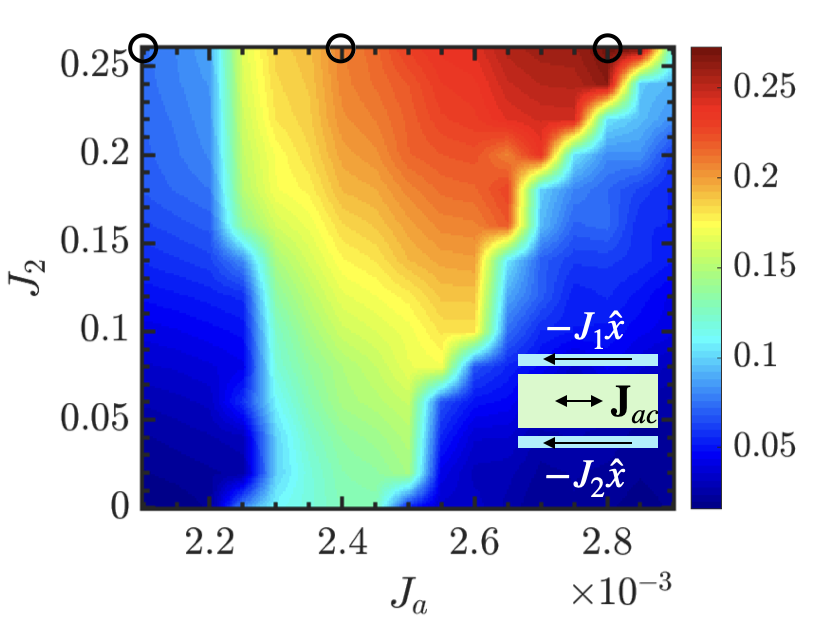}
        \caption{Phase-diagram displaying the color plot of the output voltage $V_{DC}$ (in units of $V_0=\hbar/2et_{GL}$) as a function of $J_2$ and the amplitude of the AC current (both in units of $j_0 = \sigma V_0/\xi(0)$). We fix $J_1 = 0.26$. 
        }
        \label{fig:fig6}
    \end{figure}
\begin{figure}[!t]
        \centering
        \includegraphics[width=0.9\columnwidth]{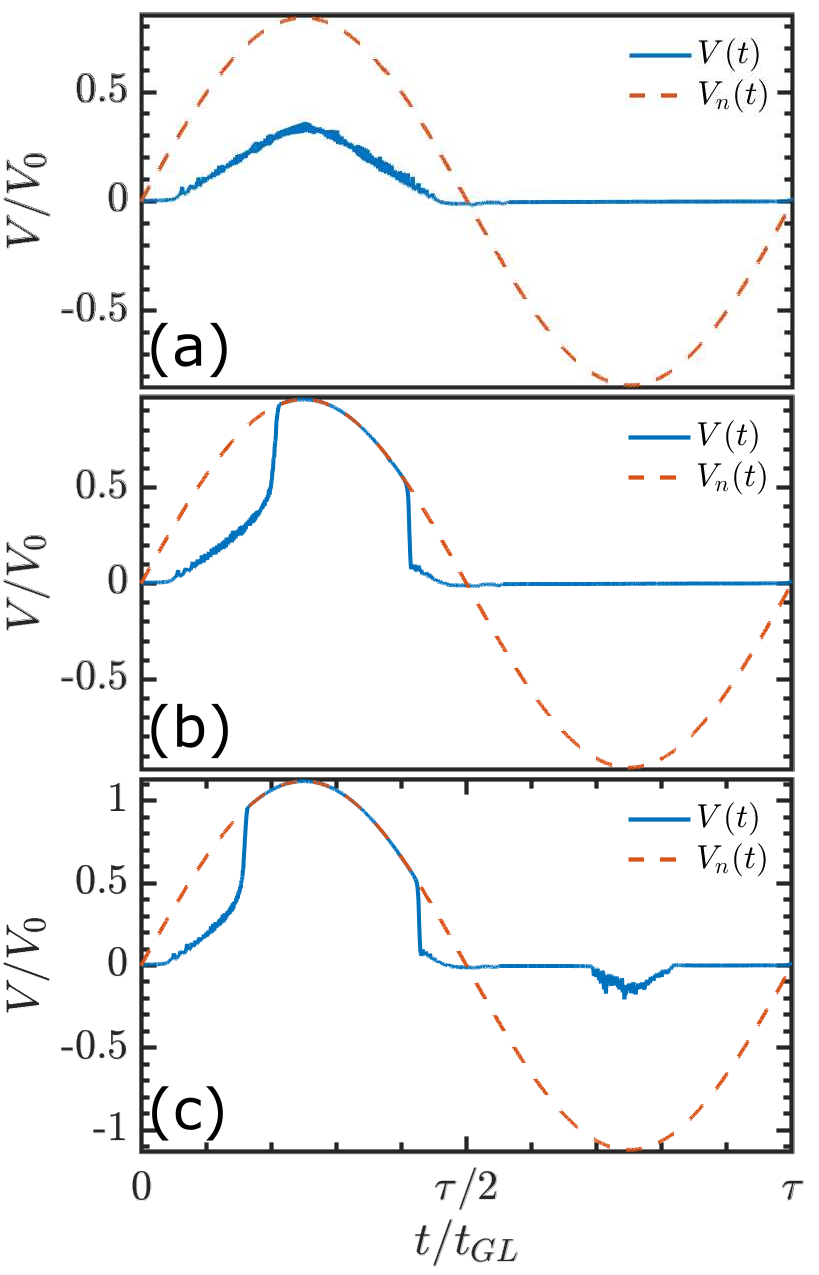}
        \caption{The voltage signal as a function of time (solid blue line) for $J_1 = J_2 = 0.26$ and three current values, $J_a = 2.10\times 10^{-3}$ (panel $(a)$), $J_a = 2.40\times 10^{-3}$ (panel $(b)$) and $J_a = 2.80\times 10^{-3}$ (panel $(c)$). In each panel, the red dashed lines represent the voltage if the system was in the normal state.}
        \label{fig:fig7}
    \end{figure}

    %
    In Fig.~\ref{fig:fig6} we show the output average voltage $V_{DC}$ as a function of $J_a$ and $J_2$, with $J_1 = 0.26$ fixed.
    As compared to the case $J_2=0$ depicted in Fig.~\ref{fig:fig2}, the large DC voltage region becomes considerably wider and the values of $V_{DC}$ 
    gets considerably higher as $J_2$ increases. In particular, for $J_2=J_1=0.26$, representing a perfect antisymmetric flux profile as illustrated in Fig.~\ref{fig:fig1}$(c)$, these enhancements are maximized: the high-$V_{DC}$ region is about three times wider and the maximum output voltage twice as high as the values observed in Fig.~\ref{fig:fig2}. To understand the reason behind the improved diode effect exhibited by the antisymmetric flux profile, in Fig.~\ref{fig:fig7} we show  
    the voltage output as a function of time for three different values of $J_a$ and $J_1 = J_2 = 0.26$. As shown in panel $(a)$, for the positive branch, shortly after $t=0$, when the magnitude of the AC current is sufficient to put in motion the vortices induced by the inhomogeneous field, the voltage becomes finite and gradually increases with current. In this case, though, the dissipation process is never large enough to totally suppress the superconducting state and we get a behaviour that resembles the positive branch of panel $(a)$ in Fig.~\ref{fig:fig3}. On the other hand, the negative current branch is significantly different from the  previous studied cases. As can be seen, after a small voltage signal corresponding to the vortex induced by the inhomogeneous field leaving the sample, the voltage becomes zero for the remaining half period of negative current. In other words, the system behaves as a half-wave rectifier.
    
    In panel $(b)$, while the negative current branch continues to present a null voltage signal, vortex creation and annihilation process is now sufficient to force the transition of the system to the normal state in the positive current branch. This is also the case for the positive current branch in panel $(c)$, which also presents a new behavior for the negative branch. Here, for a certain time interval, the AC current magnitude is large enough to overcome the surface energy barrier imposed by the inhomogeneous field and vortices begin to flow in the superconductor, inducing a finite voltage. Animations of the evolution of the order parameter for each panel in Fig.~\ref{fig:fig7} can be found as Supplementary Material.
    In these three cases, we can clearly see the reason behind the increment in $V_{DC}$, once the asymmetry between the two current polarities significantly increases in comparison to the voltage-time curves shown in Fig.~\ref{fig:fig3}. Furthermore, we note that the desired scenario for a superconducting diode is obtained for the parameters shown in panel $(b)$, once we are able to obtain a large rectified voltage while simultaneously having no dissipation for the second half of the AC current signal.

    \subsection{Hot spot dynamics}

    \begin{figure}[!t]
        \centering
        \includegraphics[width=0.99\columnwidth]{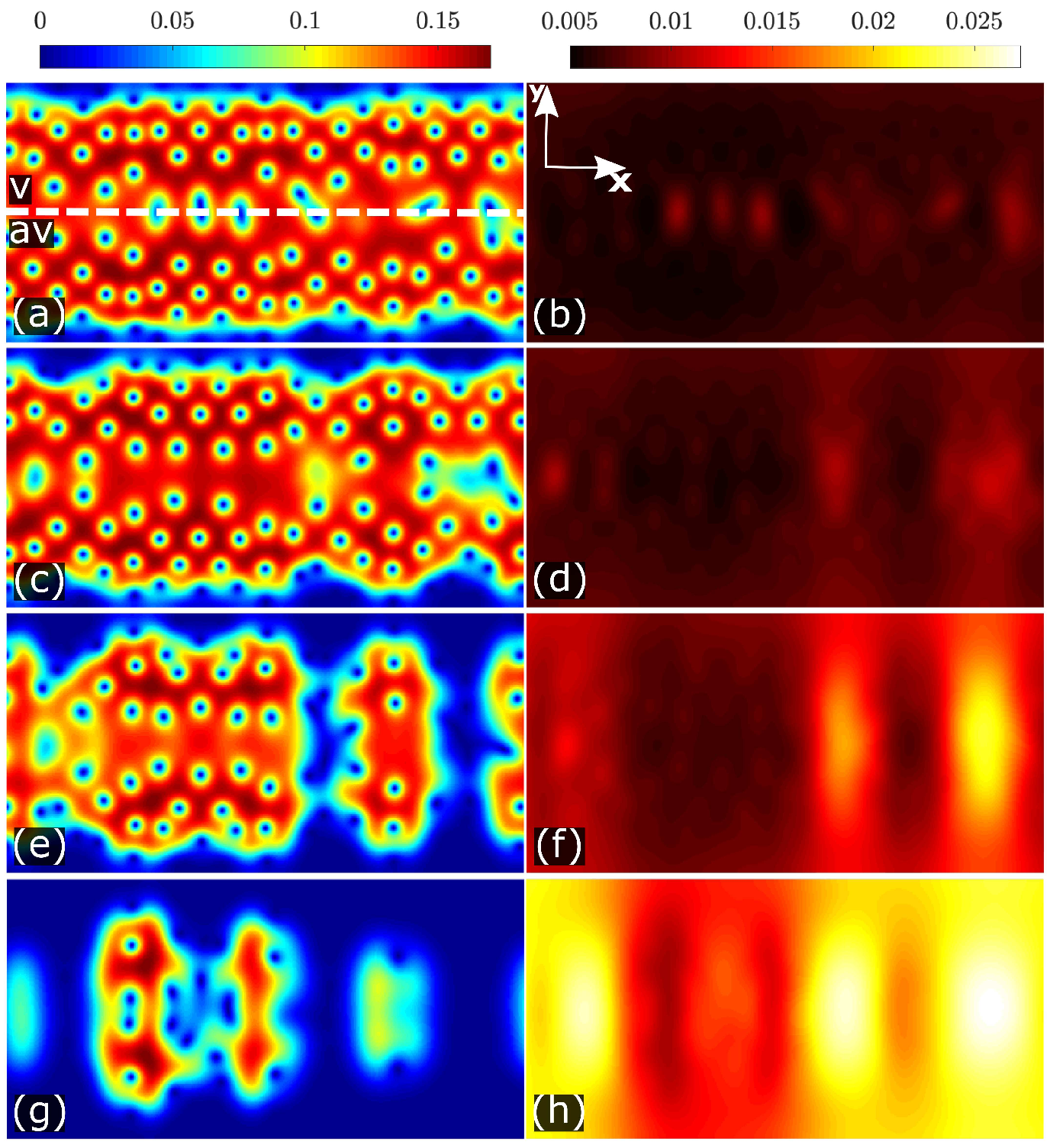}
        \caption{Evolution of the order parameter (panels in the left column) and the temperature increment $T-T_0$ (panels in the right column) before the destruction of the superconducting state for parameters of panel $(b)$ in Fig.~\ref{fig:fig7}. The white dashed line depicted in panel $(a)$ delineates the superconductor into two distinct halves, 
        with the magnetic flux being composed by vortices in the top region and antivortices in the bottom one.}
        \label{fig:fig11}
    \end{figure}
    
    To further describe our diode system, in this Subsection we detail how the vortex response to the applied current leads to the destruction of the superconducting state for positive current polarity. Fig.\ref{fig:fig11} shows the time evolution of the order parameter (left panels) and of the temperature variation $T-T_0$ near the destruction of the superconducting state for positive current polarity and parameters corresponding to panel $(b)$ of Fig.~\ref{fig:fig7}. In this figure, each panel line corresponds to the same instant of time. 
    
    From panels $(a)$ and $(b)$, we can identify that, among the processes of vortex creation, flow and annihilation, it is the latter that dissipates more heat, once each of the more pronounced hot spots seen in panel $(b)$ corresponds to a v-av collision observed in panel $(a)$. 
    Here, we can note that the perfect antissymmetric profile of the inhomogeneous field guarantees that v-av annihilation always occurs, once both objects colide at the center of the film.
    For small currents much smaller than $I_c^+$, the hot spots formed by the dissipated heat of a v-av collision gradually disappears due to the heat removal mechanism. Comparing panels $(b)$ and $(d)$, we can see an example of this in the three hot spots on the left side of panel $(b)$. In contrast, the hot spots by the right of panel $(d)$ depict the mechanism that leads to the destruction of superconductivity at $I_c^+$. In this case, heat removal is not strong enough to overcome the additional heat generated by new vortex-antivortex pairs at the same location (see panel $(c)$). As a result, the elevated temperature locally suppress the superconducting state, which attracts yet more vortex-antivortex pairs into the region. This chain reaction leads to the formation of stripes of depleted superconductivity across the width of the sample, as shown in panels $(e)$ and $(f)$. Finally, panels $(g)$ and $(h)$ shows the configuration of the system just before the total destruction of the superconducting state. As we can see, more stripes are formed along the sample and the hot spots spread throughout the superconductor.
    
    \subsection{Diode efficiency}
    
    \begin{figure}[!t]
        \centering
        \includegraphics[width=0.99\columnwidth]{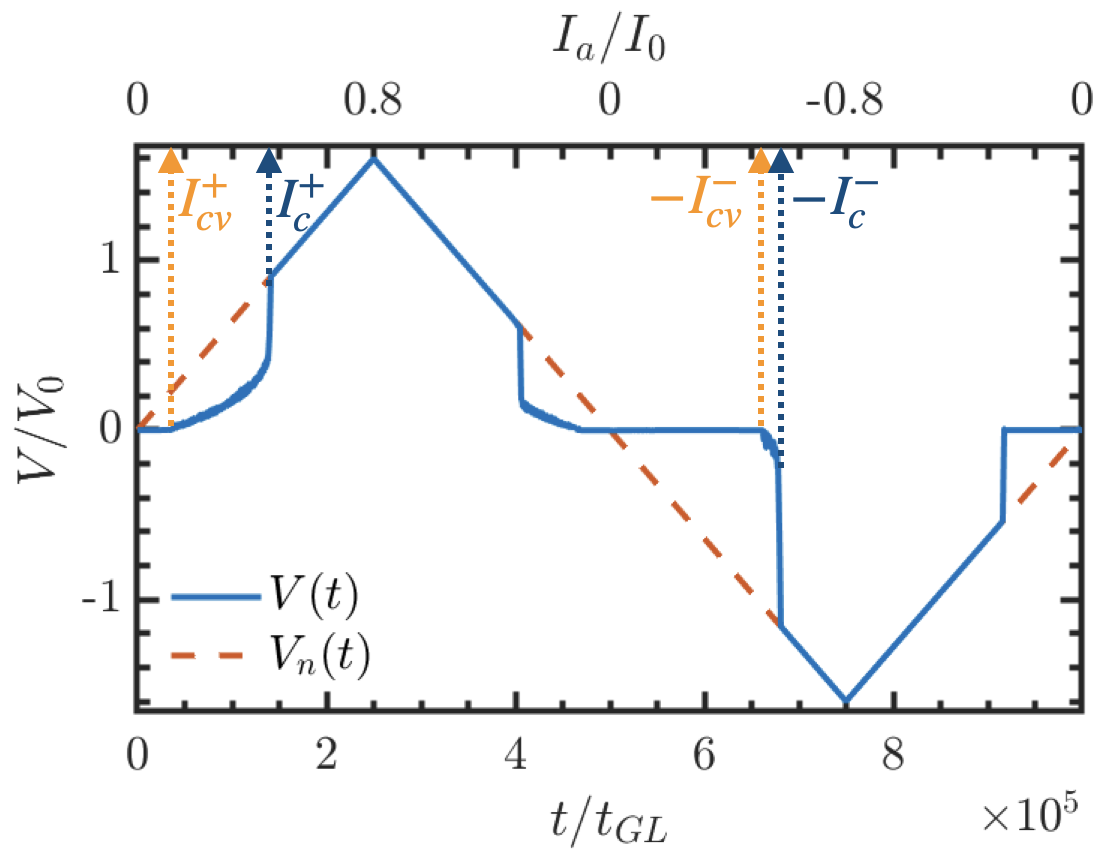}
        \caption{The voltage signal as a function of time (solid blue line) for a linear current cycle 
        with $J_1 = J_2 = 0.26$ and total sweep time $\tau = 10^6 t_{GL})$. 
        Red dashed curve represents the normal state voltage. The arrows represent the critical currents for the complete destruction of the superconducting state (dark blue) and  the onset of vortex motion (orange) at each current polarity.} 
        \label{fig:fig8}
    \end{figure}
    
    Let us now investigate the efficiency of our superconducting diode device when operating in the NSD mode, defined as $\epsilon_{NSD} = \left(I_c^--I_c^+\right)/\left(I_c^-+I_c^+\right)$, where $I_c^{\pm}$ is the critical current for the complete destruction of the superconducting state for a given current polarity. To do this, we replace the sine wave AC current by a linear current cycle, with total sweep time $\tau = 10\tau_0=10^5 t_{GL}$, thereby mimicking the usual experimental procedure of measuring the voltage in current-sweep mode. We keep $J_1 = J_2 = 0.26$ fixed, once 
    these parameters make 
    possible the occurrence of the superconducting diode scenario with largest efficiency. Fig.~\ref{fig:fig8} shows the voltage signal as a function of time for these parameters. $I_c^{\pm}$ is highlighted in this figure by the two blue arrows and their values give us $\epsilon_{NSD} = 0.124$.
    
    From Fig.~\ref{fig:fig8} we can define another useful critical current, namely the current which marks the onset of vortex motion, $I_{cv}$, which are marked by the orange arrows. 
    If we now define $\epsilon_{FFD} = \left(I_{cv}^--I_{cv}^+\right)/\left(I_{cv}^-+I_{cv}^+\right)$
    as the efficiency of the diode effect caused by the finite voltage associated with the flux-flow, we have $\epsilon_{FFD} = 0.619$ for this set of parameters. 
    The much larger value of $\epsilon_{FFD}$ 
    compared to $\epsilon_{NSD}$ can be promptly understood 
    if we notice that, 
    for positive current polarity, the profile of the self-field 
    produced by the AC current matches the profile of the external inhomogeneous field. As a result, a large amount of vortices is present in the superconductor, meaning that a small current amplitude is sufficient to start the flux-flow. In contrast, for negative current polarity, the profiles of the inhomogeneous field and the AC current self-field 
    are opposites. This means that there are no vortices in the system, unless the current amplitude is large enough to overcome the Bean-Livingston barrier.
    This is depicted at the second orange arrow in Fig.~\ref{fig:fig8} which occurs at a large current amplitude, very close to $I_c^-$. This discrepancy is the origin of the large value of $\epsilon_{FFD}$. It is worth mentioning that $I_{cv}$ is not well-defined in cases where the $VI$ characteristics is dominated by thermally activated processes. In such cases $\epsilon_{NSD}$ might be a more reliable measure of the diode efficiency.

    \begin{figure}[!t]
        \centering
        \includegraphics[width=0.9\columnwidth]{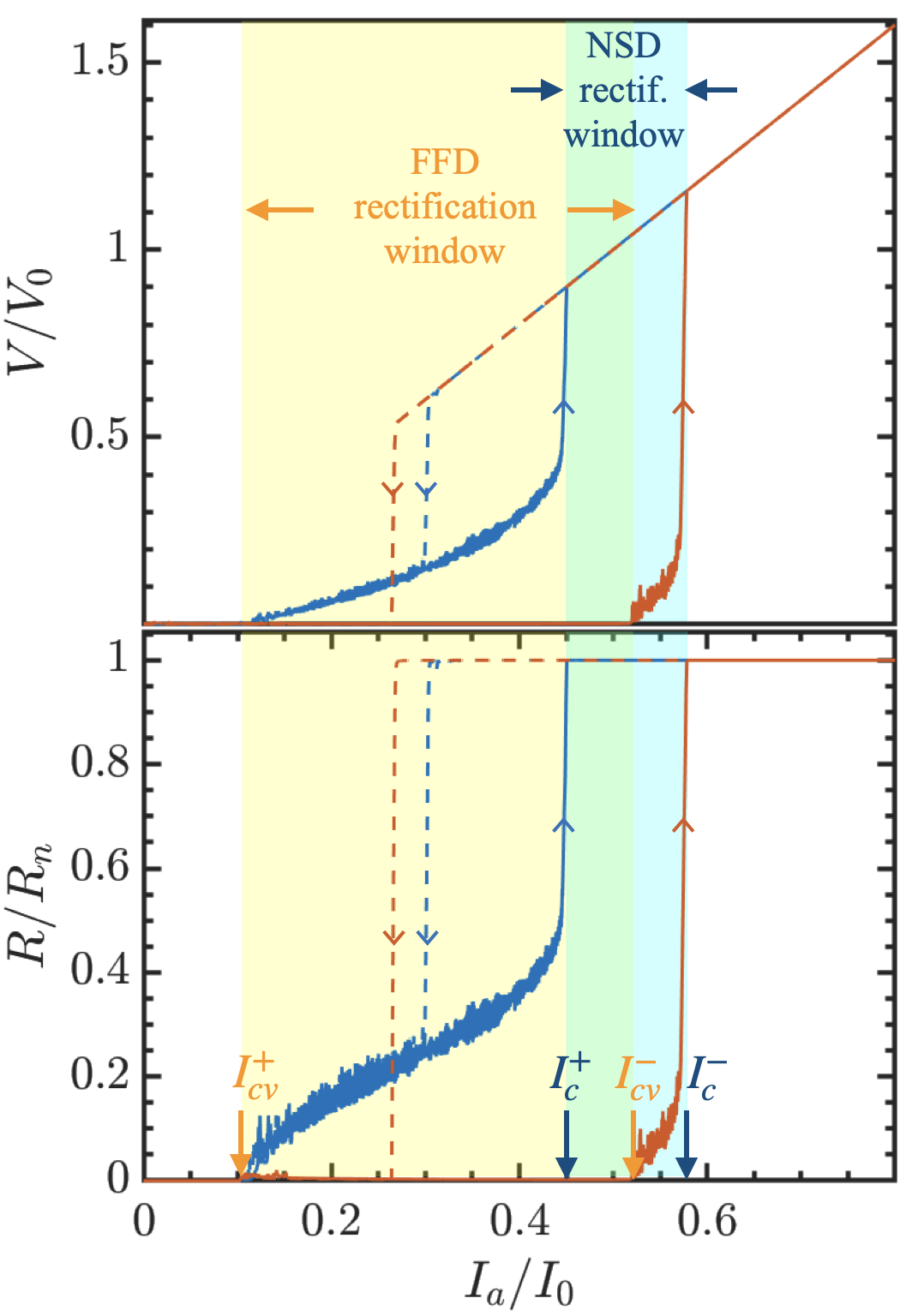}
        \caption{Top panel shows $|V(t)|$ as a function of $|I_a(t)|$. Blue and red curves represent the half period with positive and negative $J_a(t)$, respectively. Solid and dashed lines represent the regions where $|I_a(t)|$ is being increased and decreased, respectively. Yellow and blue background marks the current region of vortex and diode rectification, with green background depicting the region where they coexist. Bottom panel shows $R(t)$ as a function of $|I_a(t)|$, with the same definitions of the top panel following.}
        \label{fig:fig13}
    \end{figure}

    Fig.~\ref{fig:fig13} helps in the visualization of the vortex and diode rectification regions described above. In this figure, we show the absolute value of the voltage, $|V(t)|$, and the resistance of our system as a function of the absolute value of the total applied current, $|I_a(t)|$. 
    Regions of vortex and superconducting half-wave rectification are highlighted in the figure by yellow and blue  backgrounds, respectively, with the green background denoting the current region where they overlap. From this figure, the two key factors to improve the efficiency of a superconducting diode system becomes apparent. First, for negative current polarity, one should seek to increase the Bean-Livingston energy barrier, in such a manner as to increase $I_c^-$. Simultaneously, for the positive current polarity, the goal is to approach $I_c^+$ to the value of $I_{cv}^+$ displayed in the figure. For that, the heat induced by flux-flow and vortex-antivortex annihilations must be high enough to induce supercritical hot spots that can drive the entire system into the normal state. 
    
    \subsection{Optimizing diode efficiencies: the role of heat removal and sweep rate}
\begin{figure}[!t]
        \centering
        \includegraphics[width=0.9\columnwidth]{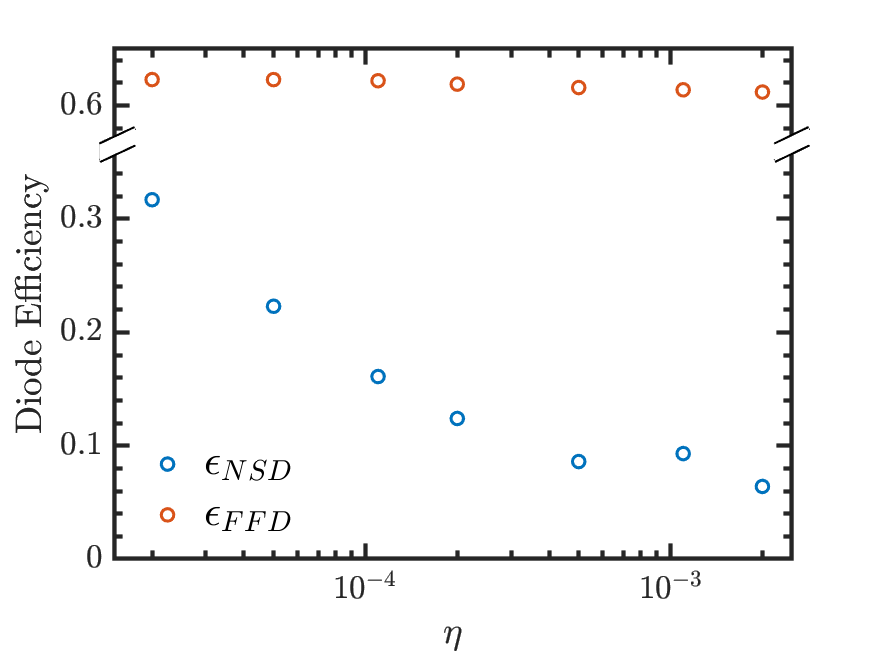}
        \caption{Efficiencies of the superconducting, $\epsilon_\text{NSD}$ (blue circles), and vortex, $\epsilon_\text{FFD}$ (red circles), diode effects as a function of the heat transfer coefficient $\eta$. Here, we set $\tau = 10^6 t_{GL}$. 
        }
        \label{fig:fig9}
    \end{figure}

\begin{figure}[!t]
        \centering
        \includegraphics[width=0.9\columnwidth]{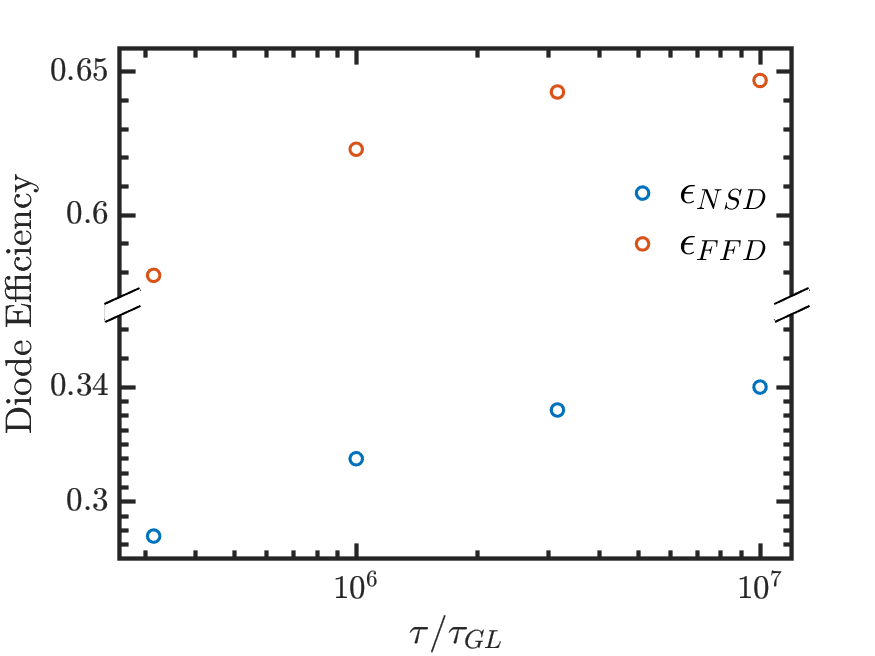}
        \caption{Efficiencies of the superconducting, $\epsilon_\text{NSD}$ (blue circles), and vortex, $\epsilon_\text{FFD}$ (red circles), diode effects as a function of total sweep time $\tau$. Here, $\eta = 2 \times 10^{-5}$.}
        \label{fig:fig10}
    \end{figure}

    So far, when solving the heat diffusion equation, we have set $\eta = 2\times10^{-4}$, which mimics a substrate of moderate heat removal efficiency. Once we have seen that the destruction of the superconducting state is governed by the emergence and expansion of hot spots through the system, it is natural to assume that the efficiency of our diode system depends on the heat removal coefficient $\eta$. Fig.~\ref{fig:fig9} shows both $\epsilon_{NSD}$ and $\epsilon_{FFD}$ as a function of $\eta$, for values raging from $2\times10^{-5}$ (weak heat removal) to $2\times10^{-3}$ (strong heat removal).
    %
    %
    The vortex diode efficiency presents a weak dependence on $\eta$, which is a result of the low heat dissipation produced by vortices when driven by currents just above $I_{cv}^+$ (see the low flux-flow resistance regime shown in Fig.~\ref{fig:fig10}). However, as the current increases above $I_{cv}^+$, the rate of v-av collisions at the equatorial line of the superconducting film increases, leading to the formation of hot spots as discussed in the previous subsection. For high $\eta$, weak hot spots are quickly removed by the substrate so that a current much larger than $I_{cv}^+$ is necessary to drive the system into the normal state. In contrast, for small $\eta$, even weak hot spots, created in the early stages of the flux-flow regime, can grow and drive the film into the normal state. This can considerably reduce the value of $I_c^+$, bringing it closer to $I_{cv}^+$. On the other hand, during the negative current branch, the reduction of $I_c^-$ induced by weak heat removal is much less pronounced because $I_c^-$ is already very close to $I_{cv}^-$, even in the strong heat removal regime. Combined, these results lead to a pronounced enhancement of $\epsilon_{NSD}$ from 0.067 to 0.317 as $\eta$ is reduced from $2\times10^{-3}$ to $2\times10^{-5}$. 
    Therefore, we can conclude that, to increase its efficiency a superconducting diode system should have the weakest possible heat removal scenario.


    It is worth mentioning that our 2D simulations are limited to the assumption that the thermal healing length is smaller than the combined thicknesses of film and substrate. In actual experiments, this condition is not generally fulfilled and typical values of $\eta$ can be much smaller than those used in our simulations~\cite{vodolazov2005}. In those cases, hot spots can nucleate as soon as flux-flow commences, thus aligning $I_c^+$ to $I_{cv}^+$ and thereby $\epsilon_{NSD}$ to $\epsilon_{FFD}$. Bearing this in mind, we speculate that the sharp superconductor-to-normal transitions observed in Ref.~\cite{hou2023} are probably triggered by hot spots induced by the fast penetration of vortices or vortex-antivortex annihilation events in a scenario of very small $\eta$.   
    
    
    Another important factor that influences the efficiency of the diode device is the total sweep time $\tau$, which determines the sweep rate along the current cycle. 
    Fig.~\ref{fig:fig10} shows both $\epsilon_{NSD}$ and $\epsilon_{FFD}$ as a function of $\tau$, with $\eta = 2\times10^{-5}$ fixed. As can be seen, both efficiencies present similar behavior, rapidly increasing for small values of $\tau$ and then approaching a constant value for large values. 
    By increasing $\tau$, we observe a decrease in both $I_c^{\pm}$ and $I_{cv}^{\pm}$. The large efficiency occurs because $I_c^+$ ($I_{cv}^+$) decreases in a larger rate than  $I_c^-$ ($I_{cv}^-$). In this case, a faster increase rate of the current, \textit{i.e.}, a lower value of $\tau$, leads to larger $I_c$ and $I_{cv}$, once there is not enough time for the system to reach the stable state for each region of current values. 
    
    \subsection{Optimizing diode efficiencies: the role of the inhomogeneous field amplitude and shape}

    \begin{figure}[tb]
        \centering
        \includegraphics[width=0.9\columnwidth]{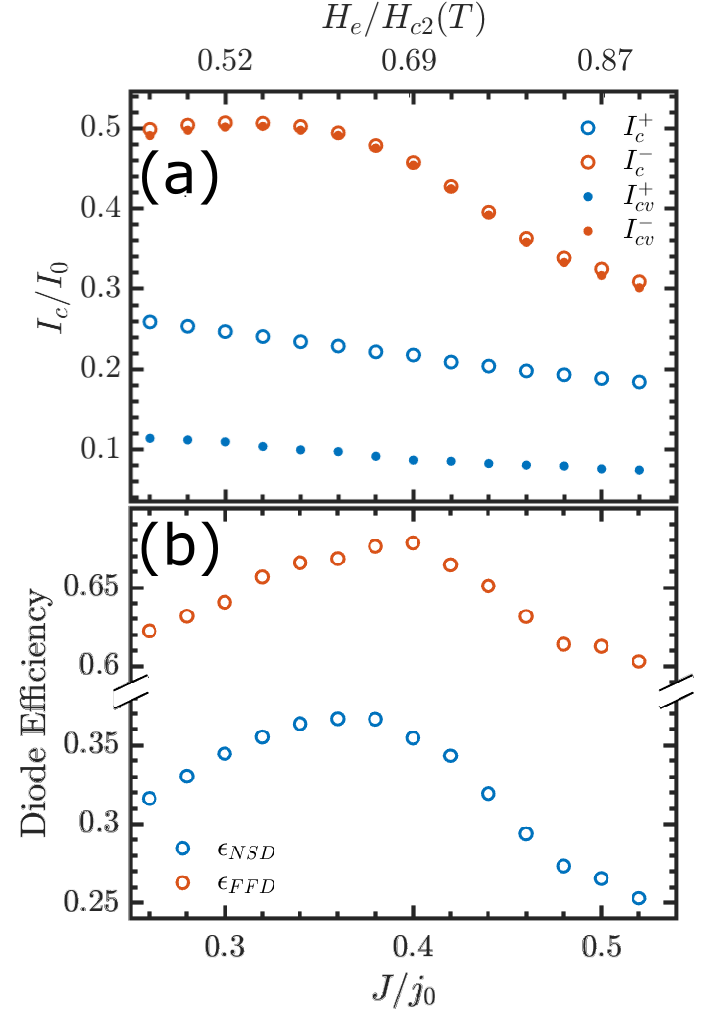}
        \caption{Panel $(a)$ shows the critical currents $I_c^+$ (blue circles), $I_c^-$ (red circles), $I_{cv}^+$ (blue dots) and $I_{cv}^-$ (red dots) as a function of the current density in the side stripes $J$. Panel $(b)$ shows $\epsilon_{NSD}$ (blue circles) and $\epsilon_{FFD}$ (red circles) as a function of $J$. Top $x$ axis relates $J$ with the value of the inhomogeneous field at the sample edges $H_e$.}
        \label{fig:fig14}
    \end{figure}
    In the above discussion, we have studied the efficiency of our diode system considering a fixed antisymmetric field profile, given by $J_1 = J_2 = 0.26$ and distance $s = 5\xi(0)$ between the superconducting film and the lateral stripes. 
    Here we investigate how the diode efficiency changes when varying the amplitude and shape of the external field profile.
    For that, we fix We start by maintaining $s = 5\xi(0)$, fixing $J_1 = J_2 = J$ and vary varying the value of $J$. We perform the calculations for a linear current cycle with sweep time $\tau = 10^6 t_{GL}$ 
    and $\eta = 2\times10^{-5}$. Panel $(a)$ of Fig.~\ref{fig:fig14} shows the critical currents $I_c^{\pm}$ (blue and red circles) and $I_{cv}^{\pm}$ (blue and red dots). 
    Both $I_{cv}^+$ and $I_c^+$ decrease monotonically with $J$, which follows from the fact that the inhomogeneous field profile adds to the self-field of the AC current in the positive current branch. Therefore, increasing $J$ contributes to increasing the number of vortices and antivortices as well as the rate of their collision, so that a lower current is required to initiate the dissipative regime. In contrast, $I_{cv}^-$ and $I_c^-$ present a non-monotonic dependence on $J$, which can be understood as follows. On one hand, increasing $J$ makes it necessary a larger value of the current magnitude to compensate the inhomogeneous field and induce the resistive state for negative current polarities. On the other, when $J$ becomes too strong, the inhomogeneous field depletes superconductivity near the edges so that lower values of the current magnitude are capable of driving the system into the normal state. 
    
    
    This discrepancy in the behavior of the system for each polarity leads to the existence of an optimal value of $J$ that gives the largest efficiency. 
    Panel $(b)$ of Fig.~\ref{fig:fig14} shows $\epsilon_{NSD}$ (blue circles) and $\epsilon_{FFD}$ (red circles) as a function of $J$. As one can see, $J \approx 0.36$ provides 
    the largest efficiency of the superconducting diode effect, while $J \approx 0.40$ is the optimal value for the vortex diode effect.

    \begin{figure}[!h]
        \centering
        \includegraphics[width=0.9\columnwidth]{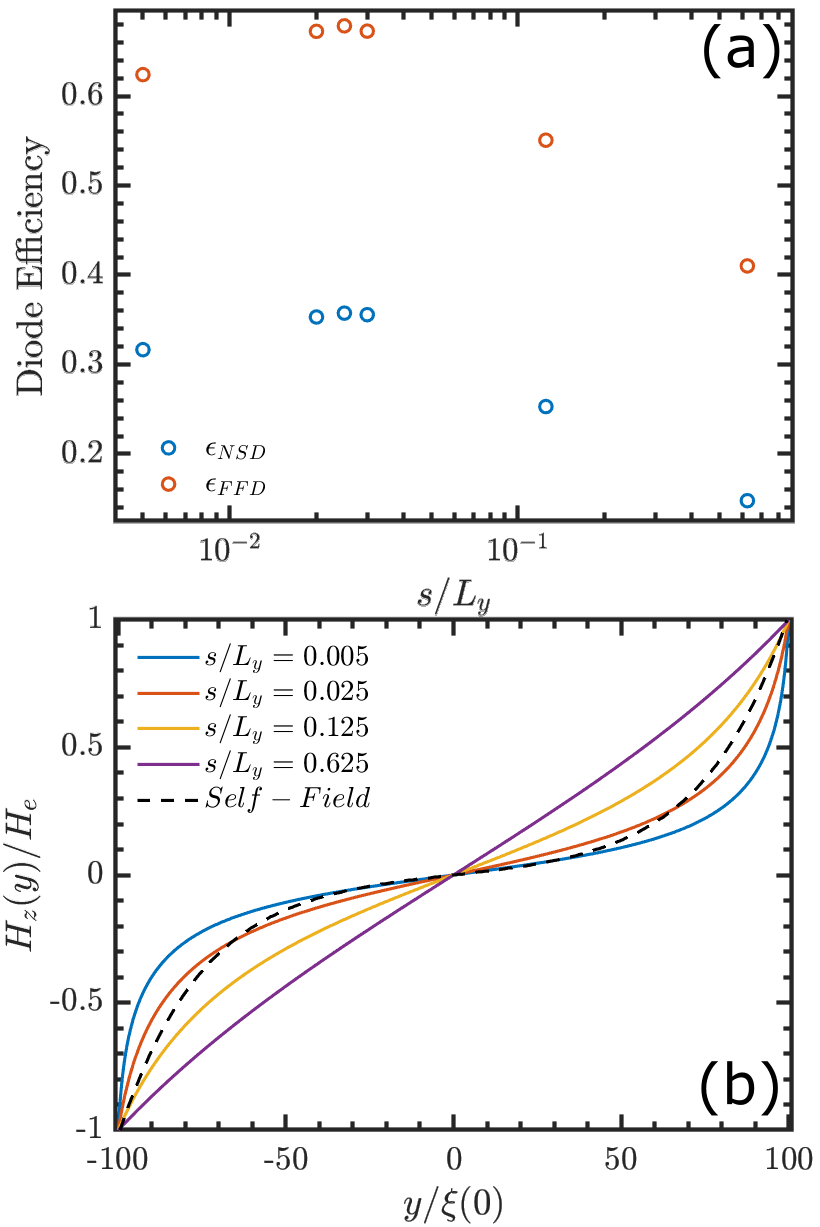}
        \caption{Panel $(a)$ shows the diode efficiency as a function of the separation between the superconducting film and the lateral wires carrying a current chosen as to induce $H_e=0.69H_{c2}(T)$. Panel $(b)$ presents the inhomogenoeus field profile for different values of $s/L_y$ (solid lines) and the self-field profile of the central superconducting film (black dashed line). All curves are normalized by the field at the edges}
        \label{fig:fig15}
    \end{figure}

    Although the previous discussion specifically deals with an inhomogeneous field generated by the side stripes currents, we note that existence of a maximum efficiency does not depend on the origin of the inhomogeneous field, but solely on its magnitude. For instance, the external field could be produced by a ferromagnetic material, as discussed in Fig.~\ref{fig:fig1}. To make our analysis more general, the top $x$ axis in Fig.~\ref{fig:fig14} relates the value of $J$ with the magnitude $H_e$ of the inhomogeneous field at both edges of the superconducting film. 
    In this case, the maximum efficiency for the superconducting diode effect occurs for $H_e \approx 0.62 H_{c2}(T)$ and for the vortex diode effect at $H_e \approx 0.69 H_{c2}(T)$. This result offers valuable guidance for achieving optimal diode efficiency experimentally in setups where the diode effect is induced by a inhomogeneous field, irrespective of its source.

    Let us now investigate how the diode efficiency depends on the shape of the inhomogeneous field. Different shapes can be induced by changing the separation $s$ between the wires and the main superconducting film: for small separations, the profile is more concentrated at the film edges, while for large separations the profile tends to a linear shape (see solid curves in panel $(b)$ of Fig.~\ref{fig:fig15}). In particular, we note that for $s/L_y = 0.025$ the profile resembles the self-field induced by the applied current in the absence of vortices. In panel $(a)$ of Fig.~\ref{fig:fig15}, we show the diode efficiencies as a function of $s/L_y$. For each value of $s$, the current density in the wires, $J$, was chosen so as to induce $H_e=0.69H_{c2}(T)$ at the film edges. It turns out that the optimum efficiency is provided precisely by $s/L_y = 0.025$. This leads to the conclusion that the most efficient way of promoting the superconducting diode effect is by using inhomogeneous field textures that replicate the self-field so as to fully compensate the effect of the applied current in a given polarity, thus keeping the device in the superconducting state.

\section{\label{sec:level4}Conclusions}
 
In conclusion, we have explored the interplay between different operation modes of the superconducting diode effect 
in thin superconducting films subjected to external asymmetric flux profiles. Our findings highlight the potential of inhomogeneous fields to create highly asymmetric conditions for vortex entrance and exit, unveiling highly efficient vortex diode effects in plain superconducting films. Notably, the diode effect observed in our study does not require sophisticated nanofabrication techniques, as the inhomogeneous field can be induced by plain ferromagnetic films underneath the superconductor or by currents injected in microsized superconducting wires coplanar with the main superconducting film. 
A centimeter-sized version of the latter setup, but using an in-plane homogeneous field, was explored long ago in~\cite{Swartz1967}, finding diode efficiencies reaching up to 67\%, similar to ours. However, it is unclear whether the origin of non-reciprocity is the same in both studies, since in our proposed micron-scale setup no external homogeneous field is required.
%

Moreover, our results demonstrate that the perfect antisymmetric flux profile represents the most efficient scenario for the diode effect. This profile maximizes the difference between the Bean-Livingstone barriers in opposite edges, leading to the highest vortex diode efficiency. Additionally, the built-in coexistence of vortices and antivortices in this scenario promotes a high rate of vortex-antivortex annihilations at the equatorial line of the sample during the positive current branch. This induces hot spots, even in the early stages of the flux-flow regime, bringing forward the transition to the normal state. In contrast, during the negative current branch, the current self-field compensates the external antisymetric field profile, eliminating vortices and antivortices from the sample, thus inducing the system into a  perfectly superconducting state. Overall, for a wide range of parameters, the system behaves as a superconducting half-wave rectifier, presenting no resistance or heat flow for half the current cycle.

Finally, we have demonstrated that the efficiency of the superconducting diode effect can be significantly enhanced by reducing the heat removal capabilities of the substrate and adjusting the sweeping rate in current sweep measurements or the frequency of the AC excitation in DC versus AC measurements. These insights not only contribute to the fundamental understanding of extrinsic diode effects in superconducting systems but also offer practical strategies for optimizing the performance of superconducting diode platforms in various applications.



\begin{acknowledgments}
We thank A. V. Silhanek for bringing Reference~\cite{Swartz1967} to our attention. LRC and ES thank the Brazilian Agency FAPESP for financial support, grant numbers 20/03947-2 and 20/10058-0, respectively. CCSS thanks financial support from Conselho Nacional de Desenvolvimento Científico e Tecnológico - Brasil (CNPq), grant number 312240/2021-0, and Universidade Federal de Pernambuco, Edital Produtividade Qualificada. This work was financed in part by Coordenação de Aperfeiçoamento de Pessoal de Nível Superior - Brasil (CAPES), Finance Code 001. 

\end{acknowledgments}



\bibliography{bibliography}

\end{document}